# Dynamical properties and search of variable stars: NGC 1960

Gireesh C. Joshi
P. P. S. V. M. I. College
Nanakmatta, U. S. Nagar-262311
09690675769
gchandra.2012@rediffmail.com

**Abstract**

*The total-to-selective extinction $R_V$ in the direction of a cluster is found to be 3.12 ± 0.2 (close to its normal value). We derive the luminosity and mass functions for the cluster main sequence stars. The mass function slope is found to be −2.29 ± 0.20 which is close to Salpeter value. We find evidence of mass segregation process in the cluster which is not yet dynamically relaxed. We have performed time series photometric observations to detect variable stars within star cluster NGC 1960. The DAOPHOT-II package is utilized to estimate the apparent stellar magnitudes of stars. The secondary standardization method is applied to the transformation of these apparent magnitudes into standard values. The magnitude-time diagrams (light curves) of stars are constructed to identify possible variability nature within them. The stars, having sufficient magnitude variation with time, are considered to be variable stars and their period values have computed through PERIOD04 package. These periodic values of variables are used to construct their corresponding phase diagrams. Here, we are reporting short periodic variables through the photometric analysis of science frames of whole night observations. Their type and variability nature, have been prescribed on the basis of information about amplitude, period and shape of phase diagrams. The location of variables on colour-magnitude-diagram is effective to constrain the history of stellar evolution. Our present analysis indicates that the variability fraction of massive stars is found to be high in the comparison of lighter members.*

## 1. Introduction

Open star clusters (OSCs) are studied to understand the evolution of galactic arms and disc as well as they have also hosted of various types of variable stars. The variation within stars are occurred due to the following reasons: (i) the accretion of surrounding matter, (ii) planet transition, (iii) the binary and triplet stellar systems, (iv) plasma transportation on the stellar surface, (v) mass of stars, (vi) environment of surrounding of stars, (vii) properties of interstellar medium, (viii) age of a star, (ix) the effect of tidal forces of nearby stars, and (x) nucleus activities in interior of star. The variable candidates were identified by inspecting their light curves [1]. The identification of variable stars is a crucial task due to following reasons: (i) the continuous observations in multi nights (each night having contained observations about 6 hours or more), (ii) the quality of observational data, (iii) the precision in the procedure of stellar magnitude estimation, (iv) the standard magnitudes of stars which depends on the transformation coefficients of the standardization night, (v) the comparative astrometry among various science frames, (vi) the procedure of cleaning (bias-subtraction, flat fielding, cosmic-ray reduction, etc.) (vii) the value of exposure time of each individual science frame, and (viii) the efficiency of procedure of detection of variability of stars (secondary standardization methodology i.e., absolute photometry or differential photometry). For more accurate and valuable period, a large amount of photometric data would be acquired on a variety of variable stars [2]. Pulsating variable stars are the most important objects due to repeating cycle of expanding and contract of the stars. These pulsating variables are distinguished by their periods of pulsation and the shapes of their light curves [3]. Their census, including pastors and binaries, can provide important clues to stellar evolution and the host star clusters [4]. The variable stars are a natural target of study for any civilization due to their correlation between period and total light output, which allowed them to become the first rung in the astronomical distance ladder [5]. Now days, Lomb–Scargle folding [6,7] and Fourier transforms [8] is prominently used for knowing the periodic nature of variance due to their better computing ability.

*Table 01: Parameters of the cluster derived in the present study along with those given in the literature.*

| Parameter | Joshi & Tyagi (2015) | Sharma et al. (2006) | Sanner et al. (2000) |
|---|---|---|---|
| Core radius (pc) | 1.6±0.3 | 1.2 | – |
| Cluster radius (pc) | 5.2±0.4 | 5.4 | – |
| Reddening(mag) | 0.23±0.02 | 0.22±0.016 | 0.25±0.02 |
| $V - M_V$ (mag) | 11.35±0.10 | 11.30 | 11.37±0.02 |
| Distance (kpc) | 1.34±0.03 | 1.33 | 1.32±0.12 |
| log Age(yrs) | 7.35±0.05 | 7.4 | 7.2±0.2 |
| Proper motion (mas/yr) | -0.08±0.11, -5.41±0.11 | – | – |

On this background, we are represented the new identified variable stars in the field of OSC 1960. Present studied cluster is widely studied by many researchers in the past and we had also been reported the structural parameters of this cluster through the statistical analysis [9]. The comparative results of the cluster *NGC 1960* is listed in Table 01. In the present work, we have been taken of advantage of our new photometric catalogue [9] for studying the dynamical behaviour of this cluster and also used in the procedure of secondary standardization. This paper is organized as follows: The total to extinction value is prescribed in Section 2. The dynamical properties have been described in Section 3. The secondary standardization method and details of time series photometric observations is listed in Section 4. The new identified variable stars and their corresponding phase diagrams are shown in Section 4. The discussion and conclusions are described in Section 5.

## 2. Extinction Law

The emitted light of the star cluster is scattered and absorbed when it passes through the interstellar dust and gas (normal reddening law is applicable in the absence of these materials [10]). The said phenomena are highly found in the blue light compare, then the red light leads to redden appearance of stars.

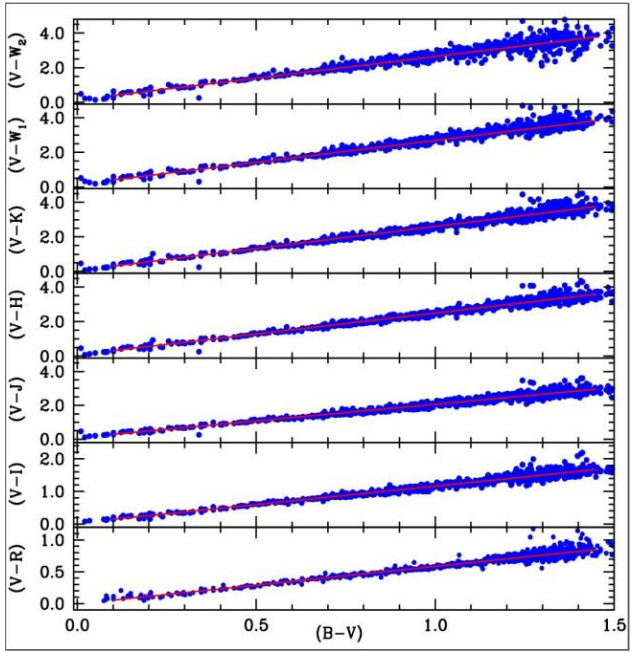

*Figure 01*: The (B−V) vs (V−λ) diagrams, where λ represents R, I, J, H, K, $W_1$, and $W_2$.

The quantity of dust and gas has been changing the nature of reddening law which has been investigated through using (V−λ)/(B−V) TCDs [11], where λ is any broadband filter, namely R, I, J, H, K, $W^1$, and $W^2$. Here, W1 and W2 are mid-IR pass-bands used for Wide-field IR Survey Explorer i.e. WISE [12]. The different (V−λ)/(B−V) TCDs of MPMs have been used to separate the influence of the extinction produced by the diffuse interstellar material from that of the intra-cluster medium [11], (TCDs are depicted in Fig. 01. The slope ($m_{cluster}$) of each TCD obtained through a best linear fit. The slope values are listed in Table 01 along with normal values and our finding are quite comparable to diffuse interstellar material. A total-to-selective extinction $R_{cluster}$ (results are listed in Table 02) is determined as [13]:

$$R_{cluster} \times m_{normal} = m_{cluster} \times R_{normal}$$

where $R_{normal}$ = 3.1 for the diffuse inter-stellar matter, which is closed to estimate mean value of $R_{cluster}$ (3.15 ± 0.08).

## 3. Dynamical study of the cluster

*The various dynamical properties are described as below:*

*(a) Luminosity and Mass functions:* The total number of cluster members in different magnitude bins refereed as the luminosity function (LFs). We considered those stars for studying the luminosity and mass functions, which has a magnitude greater than 19 mag. In present work, stellar magnitude of our photometry merged with the old deep photometry [13]. The estimated numbers of stars in each magnitude bin for the cluster (NC) are given in Table 3.

*Table 02*: The slopes of the (V−λ)/(B−V) TCDs.

| Color Value | Estimated value in the present study | Normal value | Total-to-Selective absorption |
|---|---|---|---|
| $\frac{V-R}{B-V}$ | 0.59±0.01 | 0.55 | 3.33 |
| $\frac{V-I}{B-V}$ | 1.13±0.01 | 1.10 | 3.18 |
| $\frac{V-J}{B-V}$ | 1.97±0.01 | 1.96 | 3.12 |
| $\frac{V-H}{B-V}$ | 2.42±0.01 | 2.42 | 3.10 |
| $\frac{V-K}{B-V}$ | 2.55±0.02 | 2.60 | 3.04 |
| $\frac{V-W_1}{B-V}$ | 2.53±0.02 | – | - |
| $\frac{V-W_2}{B-V}$ | 2.50±0.03 | – | - |

*Table 03*: Luminosity and Mass Functions of the stars in the (B−V)/V CMDs for the 712 MPMs of the cluster.

| V range (mag) | $N_c$ | Mass range ($M_\odot$) | $\bar{m}$ ($M_\odot$) | $log(\bar{m})$ | $log(\phi)$ |
|---|---|---|---|---|---|
| 9-10 | 3 | 7.95-6.25 | 7.15 | 0.854 | 1.460 |
| 10-11 | 7 | 6.25-4.43 | 5.32 | 0.726 | 1.670 |
| 11-12 | 8 | 4.43-3.00 | 3.85 | 0.586 | 1.675 |
| 12-13 | 20 | 3.00-2.08 | 2.48 | 0.395 | 2.099 |
| 13-14 | 36 | 2.08-1.60 | 1.85 | 0.268 | 2.500 |
| 14-15 | 44 | 1.60-1.31 | 1.46 | 0.165 | 2.705 |
| 15-16 | 81 | 1.31-1.11 | 1.24 | 0.092 | 3.052 |
| 16-17 | 153 | 1.11-0.93 | 1.03 | 0.012 | 3.298 |
| 17-18 | 131 | 0.93-0.80 | 0.87 | -0.059 | 3.302 |
| 18-19 | 184 | 0.80-0.68 | 0.75 | -0.128 | 3.417 |

The subsequent evolution of cluster may be understood through the initial mass function (IMF) but dynamical evolution constraints the direct measurement of IMF. Therefore, MF can be introduced, which can be the relative numbers of stars per unit mass and the corresponding

expression is expressed as
$$N(\log M) \propto M^{\Gamma}.$$
The slope, $\Gamma$, is defined as
$$\Gamma = \{d \log N (\log m)\}/\{d \log m\}$$
$N \log(m)$ is the number of stars per unit logarithmic mass. The masses of MPMs can be determined by comparing their observed magnitudes with those predicted by a stellar evolutionary model (For known age, reddening, distance and metallicity). The MF values are given in Table 2 by considering the negligible effect of field star contamination. Fig. 2 shows the MF in the cluster fitted for the MS stars with masses $0.57 \leq M/M_\odot < 7.95$. The MF slope ($\Gamma$) for cluster comes out to be $-2.29 \pm 0.20$ (error computed through Poisson Statistics) which is in close agreement with the Salpeter MF slope of -2.35 (Salpeter 1955) within the given uncertainty, however, stepper than the MF slope, $\Gamma = -1.80 \pm 0.14$ [14].

*(b) Mass Segregation:* In this process, the stellar encounters are gradually increased, so that the higher-mass members gradually sink towards the cluster center by transferring their kinetic energy to low-mass members [15]. So, the spatial mass distribution in the cluster changes with time.

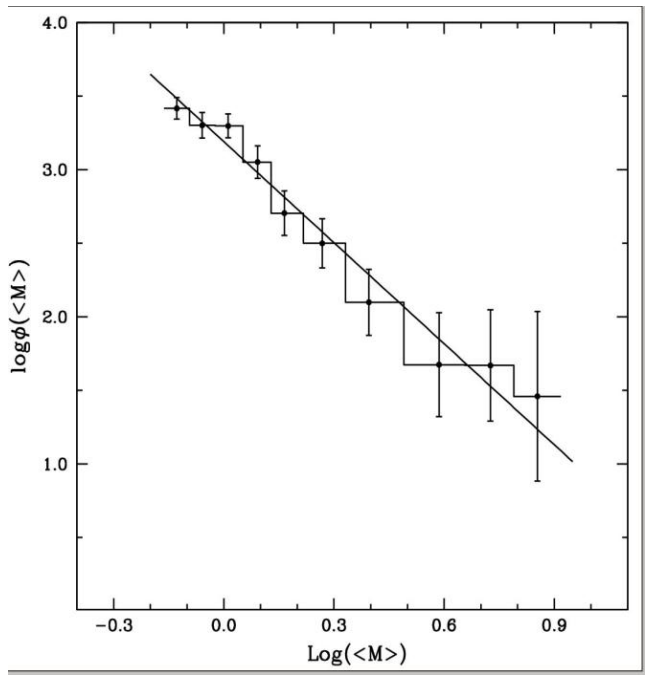

*Figure 02:* MF derived for the cluster region. The error bars represent $1/\sqrt{N}$ errors.

The radial variation of the MF slope is depicted in Fig. 3 to investigate the dynamical evolution and mass segregation process. The steepness property of the MF slope is increased with the radial distance of the cluster, which is indication of the mass segregation within the cluster. For more detail, the variation of mean mass of MPMs along the radial distance (shown in Fig. 4) indicated that the higher and lower mass stars are mainly distributed to the core and outer region of the cluster respectively, and significant

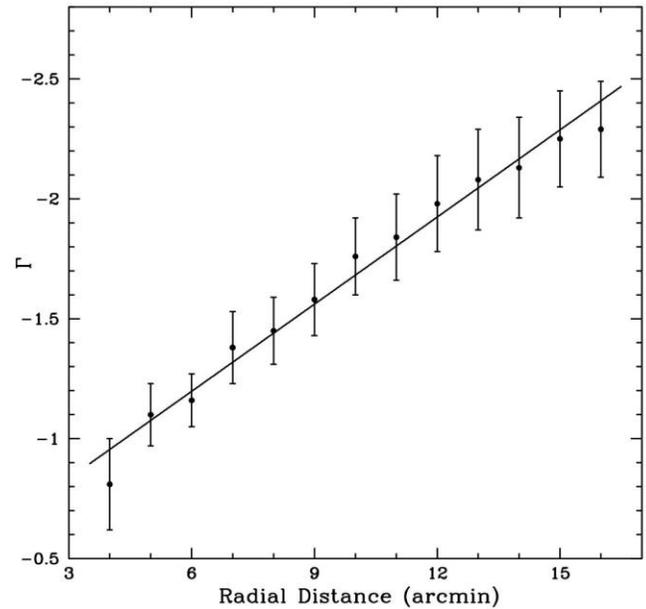

fraction of total cluster mass contained significant fraction of lower mass stars.

*Figure 03:* The variation of MF slope in the radial direction.

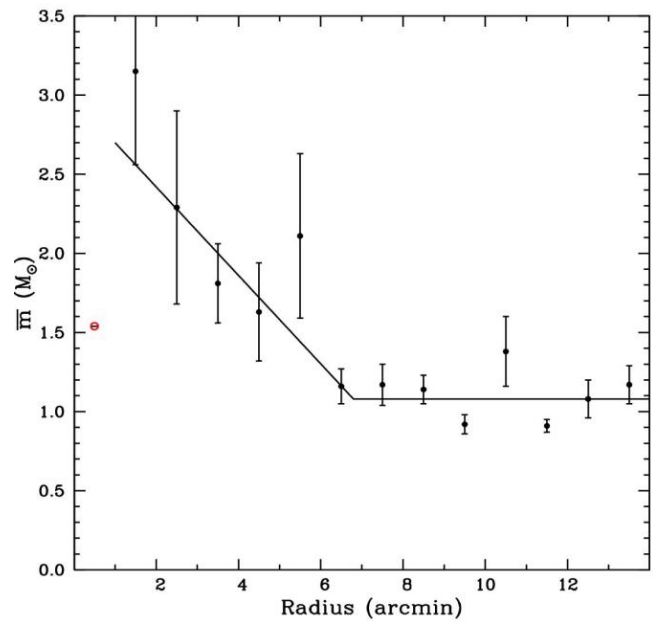

*Figure 04:* The Mass distribution of stars in the radial direction.

The dynamical relaxation time, $T_E$, is the time in which individual stars in the cluster exchange energies and their

velocity distribution approaches Maxwellian equilibrium. It can be expressed as

$$T_E = \frac{8.9 \times 10^5 (NR_h^3/\bar{m})^{1/2}}{\log(0.4N)}$$

where $T_E$ is in Myr, N is the total number of cluster members, $R_h$ is the radius (in parsecs) containing half of the cluster mass and $\bar{m}$ is mean mass of the cluster members in solar units. The mass range of MPMs found as $0.68 \leq M/M_\odot < 7.95$, which gives the total mass of the cluster as ~ 802 $M_\odot$ with an average mass of ~ 1.13$M_\odot$ per star. The contribution of the low-mass stellar population is critical for constraining the total cluster mass, which is crucial in understanding the dynamical evolution and the long-term survival of a cluster (e.g., de Grijs & Parmentier 2007). A half-mass radius of the cluster found to be 3.72 pc which is ~ 70% of the cluster radius and plays an important role in the determination of the dynamical relaxation time. Its larger value than the half radius of the cluster is a clue of deficiency of massive stars in the core, which may be occurred either undetected massive stars due to saturation limit or an ongoing mass segregation process in the cluster. The dynamical relaxation time, $T_E$, is found to be 61.9 Myr, which is higher than the cluster age (~22 Myr). This result comes due to limiting boundary of photometric criteria.

Sharma et al. (2008) [14] report a significantly different values for the total mass, mean mass and the dynamical relaxation time through 232 members, which are found by them through the statistical subtraction i.e. by subtracting the contribution of field stars from the cluster field for each magnitude bin. This method is avoided in the present study.

## 4. Secondary Standardization Method and Quality of the Data

Secondary standardization method is a technique to compute the absolute photometric magnitude of observed magnitude of stars of cluster through a dataset of standardized night of cluster (reference catalogue). This method may be effective to We have been using the secondary methodology to reduce atmospheric-effect/ estimation-error. It is purposed to take a dataset of common stars between detected stars in a science frame and reference frame. These common stars are used to find out a linear solution for transforming the observed stellar magnitudes into absolute magnitudes. Furthermore, above said linear solution has been applied for all detected stars. Here, the new UBVRI photometric catalogue [9] is used to compute the standard stellar magnitudes from the observational data as assuming the reference catalogue. The said observational data has been observed by GCJ from 1.04m Sampurnand telescope of ARIES. The whole night observations were carried out for determining the short period stars. There are two datasets for the present analysis. The first one given by Dr. Santosh Joshi (ARIES) to GCJ but the data quality is not good which indicates that either seeing is not good or observational data is highly influenced by the passing clouds. The given date night is 24 January 2012 (70 frames, 3.5 hours). The second dataset of this cluster is observed by GCJ through 1.04m Sampurnanand telescope of ARIES (Manora Peak, Nainital) on the following dates: 11 December 2013 (150 frames, 5.4 hours), 20 December 2013 (90 frames, 7.6 hours), 12 January 2015 (200 frames, 7.2 hours) and 08 Fer-bury 2015 (143 frames, 5.6 hours). This latter dataset is also not found to be excellent for searching variable stars within the cluster, but it is better than the data given by Dr. Santosh Joshi, the old dataset is utilized for checking the variable nature of variables and also used for estimation of the period but mostly avoided for small variation of stellar magnitudes. The new dataset (gathered by GCJ) was influenced by passing clouds after half observations of a night and variation of stellar magnitude is also possible due to the higher declination of the target cluster from zenith. All these datasets are gathered in V-photometric band through 1.04m Sampurnanand telescope of ARIES. The said telescope contained 2k×2k charge couple device (CCD) camera which covers 15×15 arcmin2 field of view of the target object. This observed region is a small portion of the whole cluster. The coordinates of detecting stars are found in pixel which transforms into RA and DEC through astrometry. Furthermore, the bias correction and flat-fielding of observed science frames have been completed through those bias frames and flat frames, which were observed on the corresponding night of observations. In the absence of these prescribed frames, we were utilized these frames from the nearby observed night. The exposure time of these observations is varied from 5 Sec to 60 Sec. We have been detected different number of stars in different science frames of the object which is happening due to different exposure time and other observational condition during the night.

## 5. Search of variables within the cluster

The variation of stellar magnitude with time is known as light curve. If the variation of the absolute magnitude of stars more than 3-σ of its mean value, then it is considered to be variable star. Due to the bad seeing and passing clouds, the absolute optometry does not suffice to identify the exact variable curve of new identified variation. Therefore, we are purposed to differential photometry over absolute photometry and this resultant procedure may be defined as differential-absolute photometry. The effective reduction of influenced by sky variation of observational night is the major advantage of this new purposed procedure. This sky variation is a resultant effect of passing clouds and environmental changes during observations (humidity, wind flow, declination of the target object from the zenith etc.). It is a technical fact that the comparison stars are a basic requirement of differential photometry. The practice of chosen of a single comparison star may be avoidable due to fact that the exact confirmation of the variable curve of a star is not possible due to unavailable information of occurrence of a variable curve. Therefore, we have been selected comparison stars more than for every possible variable star. These selected comparison stars are contained in the same order of photometric magnitude of V-band and reddening. The magnitude of comparison star of each epoch has been subtracted from the corresponding magnitude of the variable. After investigating these resultant light curves of variable through comparison, we have been selected smoother light curve for computing the period of variable.

*Table 04*: *The period value of variables within NGC 1960. The first column shows the ID of variables. The second and third columns represent RA and DEC respectively. The fourth and fifth columns indicate the V-magnitude and reddening of new identified variables. The last column listed the value of the period.*

| ID | RA | DEC | V | (B−V) | Period |
|---|---|---|---|---|---|
| 1123 | 05 : 36 : 28.54 | 34 : 09 : 05.8 | 15.060 | 0.743 | 0.15286 |
| 1199 | 05 : 35 : 53.49 | 34 : 08 : 09.6 | 15.155 | 1.512 | 0.16024 |
| 1350 | 05 : 36 : 21.20 | 34 : 05 : 25.4 | 15.345 | 0.766 | 0.29911 |
| 1470 | 05 : 36 : 17.96 | 34 : 05 : 38.7 | 15.497 | 0.938 | 0.34959 |
| 1549 | 05 : 36 : 39.17 | 34 : 11 : 42.8 | 15.592 | 0.778 | 0.21316 |
| 1601 | 05 : 36 : 17.84 | 34 : 06 : 31.8 | 15.668 | 0.966 | 0.63697 |
| 1701 | 05 : 36 : 36.95 | 34 : 11 : 57.0 | 15.768 | 1.184 | 0.22273 |
| 2198 | 05 : 36 : 21.81 | 34 : 05 : 34.1 | 16.197 | 0.809 | 0.25658 |
| 2323 | 05 : 35 : 44.69 | 34 : 03 : 03.4 | 16.279 | 0.984 | 0.36091 |

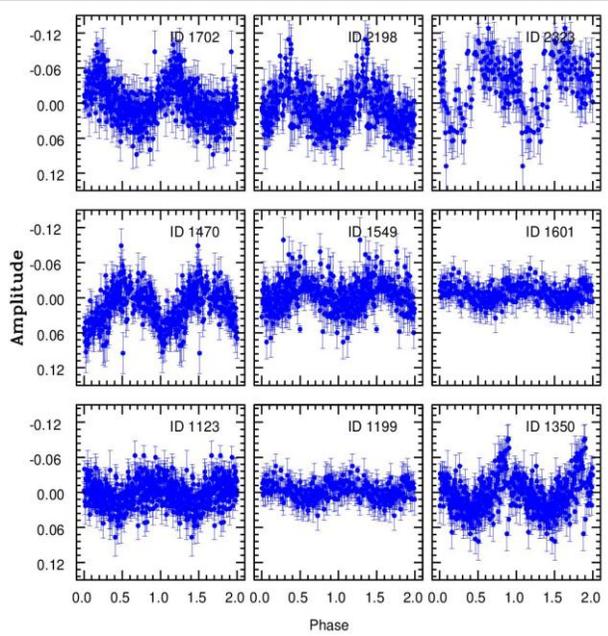

*Figure 05*: *The panels represent the light-folded-curve or phase diagram of identified variable stars within the cluster.*

The summation of scattering of variable and comparison star is the major disadvantage of differentiate photometry. To reduce the influence of scattering, we are adopting the average moving procedure [18] and applied to the time-series photometric data. However, it is noticeable fact that the amplitude of the variable curve decreases with the every completed cycle of this procedure. As a result, prescribed cycle of average-moving-procedure is repeated one/two times. The smoothness of the phase-folded diagram increases by this method. As a result, this whole procedure seems to be more reliable to evaluate the variable nature of stars. Due to the varying sky conditions, the changes in amplitude of magnitude variation are highly scattered. This scattering is highly occurring in brighter and fainter stars; therefore we have excluded the brighter and fainter stars of the cluster for searching the variable stars. Here, we have been carried out variable work for those stars which having V-magnitude between 15 mag and 16.5 mag. A total of nine variables is found in this magnitude-range. All these variables are short periodic variable stars and their period is less than a day. Their phase-folded curves are shown in Figure 5. The 'PERIOD-04' software has been utilized to estimate the period of new identified variable stars.

## 6. Conclusion

The slope of MF of main sequence (MS) seems to be −2.29 ± 0.20, the mass of prescribe MS is obtained in the range $0.68 \leq M/M_\odot < 7.95$. The present analysis indicates that steepness of MF slope increases with the radial distance of the cluster. This fact becomes as evidence of the mass segregation within the cluster. The relaxation time of the cluster has seemed to be much higher than its age due to the restricted boundary limit of red and blue sequences of photometric criteria, the present results are imprinted for denying the possibility of the dynamic relaxation of the cluster. Here, we are realizing that there is an effective improvement needed for membership determination.

The quality of observed data is poor for searching variable stars within the cluster. The present results of period are highly influenced by the observational conditions. The scattering is found to be high and our present procedure of reduction of scattering also reduces the amplitude of variable curve. Therefore, our present analysis shows further improvement to overcome the problem of amplitude reduction. Due to lack of qualitative data, it would be highly recommended that the photometric time-series data is required for accurate estimation of period of identifying variable. The poor quality of data may be reduced the possibility of finding variable stars within the cluster. As a result, present computed period values may be altered due to either good quality of new gathered observational data or required improvement of the magnitude estimation methodology of the detected stars.

## 7. Acknowledgements

facilities. Their kind help encourage me to search the literature and develop scientific improvement of this paper. GCJ is also acknowledged to ARIES, Nainital for providing observing facility duration Oct, 2012 to April 2015 and also acknowledge to Dr. Santosh Joshi with Dr. Brajesh Kumar for providing data of NGC 1960 on the night of 30 Nov, 2010, which was used to compute the absolute magnitudes of stars in various photometric bands and in the present paper, it is taken as a reference dataset.